\documentclass[a4paper]{article}
\usepackage{spconf,amsmath,graphicx,cite}

\title{Power control with partial observation in wireless ad hoc networks}

\name
  {Sara Berri $^{*,\dagger}$, Samson Lasaulce  $^\dagger$, and Mohammed Said Radjef $^*$}
	\address{$^*$ Research Unit LaMOS (Modeling and Optimization of Systems), Faculty of Exact Sciences,\\ University of Bejaia, Bejaia, 06000, Algeria \\
$^\dagger$  L2S (CNRS-CentraleSupelec-Univ. Paris Sud), Gif-sur-Yvette, France }

\usepackage[mathcal]{euscript}
\usepackage{amssymb,amsmath}
\usepackage{graphicx}
\usepackage{tikz}
\usetikzlibrary{arrows,backgrounds,snakes}
\usepackage{amsmath,amstext,amsfonts,amssymb}
\usepackage{amsbsy}

\usepackage{color}
\definecolor{darkred}{rgb}{0.5,0.0,0.0}
\definecolor{darkblue}{rgb}{0.2,0.1,0.6}

\begin{document}

\maketitle

\begin{abstract}
In this paper, the well-known forwarder's dilemma is generalized by accounting for the presence of link quality fluctuations; the forwarder's dilemma is a four-node interaction model with two source nodes and two destination nodes. It is known to be very useful to study ad hoc networks. To characterize the long-term utility region when the source nodes have to control their power with partial channel state information (CSI), we resort to a recent result in Shannon theory. It is shown how to exploit this theoretical result to find the long-term utility region and determine good power control policies. This region is of prime importance since it provides the best performance possible for a given knowledge at the nodes. Numerical results provide several new insights into the repeated forwarder's dilemma power control problem; for instance, the knowledge of global CSI only brings a marginal performance improvement with respect to the local CSI case.
\end{abstract}

\section{Introduction}
\label{sec:introduction}

\begin{figure}[h!]
\begin{center}
\begin{tikzpicture}
\node (12) at (3,0.8) {${g_1}$}; 
\node  (23) at (5,0.8) {$g'_2$};
\node  (21) at (3,-0.8) {$g_2$};
\node  (14) at (1.2,-0.8) {$g'_1$};
\node  (4)  at (0,0) {4};
\node  (1)  at (2,0) {1};
\node  (2)  at (4,0) {2};
\node  (3)  at (6,0) {3};
\draw [very thick](0,0) circle (0.3cm);
\draw [very thick](2,0) circle (0.3cm);
\draw [very thick](4,0) circle (0.3cm);
\draw [very thick](6,0) circle (0.3cm);
\draw[very thick][->,>=latex] (1) to[bend left=45] (2);
\draw[very thick][->,>=latex] (2) to[bend left=45] (3);
\draw[very thick][->,>=latex] (2) to[bend right=-45] (1);
\draw[very thick][->,>=latex] (1) to[bend right=-45] (4);
\end{tikzpicture}
\end{center}
\caption{The figure represents the studied scenario, each node $i\in\{1,2\}$ uses the power couple $(p_i,p'_i)$. $p_i$ for the own packets, $p'_i$ for the other nodes' packets.}\label{fig1}
\end{figure}
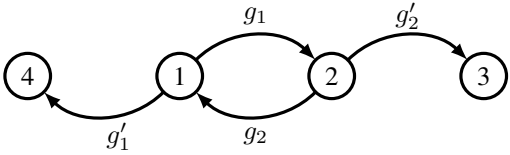

In wireless ad hoc networks nodes are typically interdependent. One node needs the assistance of neighboring nodes to relay the information it wants to send to the receiver(s). Therefore, nodes are in the situation where they have to relay signals or packets but have at the same time to manage the energy they spend for helping other nodes. To study the tradeoff between a cooperative behavior, which is necessary to convey information through an ad hoc network, and a selfish behavior which aims at managing the node energy, the authors of \cite{hub} proposed a very simple but efficient model. Their modeling has been found to be very important and insightful in the literature of ad hoc networks, as advocated by the many papers where it is exploited. The model consists in studying, in possibly large ad hoc networks, the local interaction between four nodes (see Fig. \ref{fig1}). Node $1$ (resp. $2$) wants to send information to Node $3$ (resp. $4$) and, for this purpose needs the assistance of Node $2$  (resp. $1$). In the original model of \cite{hub}, Node $1$ (resp. $2$) has two possible choices namely, forward or drop the packets it receives from Node $2$ (resp. $1$). Assuming that each node wants to maximize a utility function which consists of the addition of a data rate term (which is maximized when the other node forwards its packets) and an energy term (which is maximized when the node does not forward the packets of the other node). At the Nash equilibrium of the corresponding strategic-form game (called the forwarder's dilemma in the corresponding literature), nodes don't transmit at all. To avoid this, cooperation has to be stimulated e.g., by studying the repeated interaction between the nodes \cite{hub} \cite{c2i7} or by implementing incentive mechanisms \cite{c2i9} \cite{c2i1}. While providing an efficient solution, these models still have some limitations. The purpose of this paper is precisely to overcome those limitations.

If we interpret the model of \cite{hub} as a power control problem for which each node has to decide to transmit at maximum or minimum power to relay the packet of the other node, four limitations appear in the formulation of \cite{hub}. First, the node only chooses the cooperation power while in an ad hoc wireless network such as sensor networks, it has also to choose the power used to send its own packet. Second, the power control scheme does not take into account the quality of link between the transmitting node and the receiving node. Third, no framework is provided to tackle the scenario where the nodes have only partial observation of the actions of the other nodes and the channel gains. Fourth, no framework is provided to be able to reach any point of the feasible utility region, especially in the presence of partial observation.

The purpose of the present paper is precisely to contribute to overcoming these limitations. To this end, we exploit the recent Shannon-theoretic works \cite{benja},\cite{larrousse}. These works give coding theorems which we exploit here in a constructive manner. This allows us to find the feasible utility region of the considered problem in the presence of partial observation and, more importantly, to provide a numerical algorithm to obtain globally efficient power control policies.

The paper is structured as follows. In Sec. \ref{sec:problem-formulation}, we provide a local interaction model between four nodes of an (possibly large) ad hoc network which generalizes the famous model introduced in \cite{hub}. In Sec. \ref{sec:utility-region} we explain how the feasible utility region of a strategic-game with partial observation can be obtained and provide a numerical technique to determine decision functions; it has to be noted that this works for any utility function and not only for those assumed in Sec. \ref{sec:problem-formulation}. Sec. \ref{sec:numerical-perf-analysis} corresponds to the numerical performance analysis.

\section{Proposed problem formulation}
\label{sec:problem-formulation}

In this section, we provide a model which generalizes the model of \cite{hub} (see Fig. \ref{fig1}). The model is as follows. The quality of the four links between the four nodes is assumed to fluctuate over time. More precisely, we assume a block-fading law for the four channel gains  (see Fig. \ref{fig1}) namely, each channel gain is assumed to be constant over the duration of a block or packet and varies from block to block in an i.i.d. manner. Nodes $1$ and $2$ are the source nodes while Nodes $3$ and $4$ are the destination nodes. For each block, Node $i$, $i\in \{1,2\}$, has to choose $p_i$ the power it uses to send its own packet to the closest node and $p_i'$ the cooperative power it uses to relay the packet of the other source node. The channel gains $g_1$, $g_1'$, $g_2$, $g_2'$ and the transmit powers $p_1$, $p_1'$, $p_2$, $p_2'$ are assumed to lie in discrete sets: $\mathcal{G}_{i}=\{g^{1}_{i},\ldots,g^{N}_{i}\}$ and $\mathcal{G'}_{i}=\{g'^{1}_{i},\ldots,g'^{N}_{i}\}$, with $g^{1}_{i}=g^{\min}_{i}$, $g^{N}_{i}=g^{\max}_{i}$, $g'^{1}_{i}=g'^{\min}_{i}$ and $g'^{N}_{i}=g'^{\max}_{i}$, $N\geq1$; $\mathcal{P}_{i} =\mathcal{P}_{i}' =\{p^{1}_{i},\ldots,p^{M}_{i}\}$, with $p^{1}_{i}=P^{\min}_{i}$ and $p^{M}_{i}=P^{\max}_{i}$, $M\geq1$. Assuming that the sets are discrete is of practical interest, as there exist wireless communication standards in which the power
can only be decreased or increased by step and in which quantized wireless channel state information is used (see e.g., \cite{sesia-book-2009} \cite{david}).

The performance metric, which will be called utility, of Node $i \in \{1,2\}$ is chosen as follows:
  \begin{equation}\label{eq51}
   u_{i}(x_0,a_{1},a_{2})=\varphi(\mathrm{SNR}_{i})-\alpha(p_i+p'_i).
\end{equation}
where: $x_0=(g_1,g'_1,g_2,g'_2)$ is the \textit{global channel state}; $a_i=(p_i,p'_i)$ is the action of Node $i$; $\mathrm{SNR}_i$ is the signal to interference ratio. Since a source node sends packets through an intermediate relay node, using a two-hop communication link, $\mathrm{SNR}_i$ is defined according to the source node and relay node;
\begin{equation}
\mathrm{SNR}_i=\frac{p_{i} g_{i}p'_{-i} g'_{-i}}{\sigma^{2}},
\end{equation}
$\sigma^{2}$ being the noise variance and $-i$ standing for the other source node. The function $\varphi$ is a communication efficiency function which is assumed to be increasing and lie in $[0,1]$; it may typically represent the packet success rate; a typical choice for $\varphi$ is for example, $\varphi(x) = (1-e^{-x})^L$, $L$ being the number of symbols per packet (see e.g., \cite{goodman-pcom-2000}\cite{meshkati-jsac-2006}\cite{lasaulce-twc-2009}) or $\varphi(x) = e^{-\frac{c}{x}}$ with $c=2^r -1$, $r$ being the spectral efficiency in bit/s per Hz \cite{belmega-tsp-2011}. The parameter $\alpha \geq 0$ allows one to assign more or less importance to the energy consumption of the node. Note that the assumed expression of the SNR may for instance, be justified when nodes implement the amplify-and-forward protocol to relay the signals or packets to the destination \cite{elgamal-book-2011}. The set of nodes $\mathcal{I}=\{1,2\}$, the action sets $\mathcal{A}_i = \mathcal{P}_i \times \mathcal{P}_i'$, and the utility functions $u_1$, $u_2$ define a \textit{static or one-shot strategic-form game.} The original one-shot game model of \cite{hub} can be obtained by assuming that $\varphi$ is a step function, $p_i$ is constant, $p_i'$ is binary, and all the channel gains are constant.

The problem we want to solve in this paper is as follows. Assuming that the nodes interact over $T\geq 1$ stages or blocks and that they have a certain knowledge of the channel state, what is the utility region of the system? And which power control policy should be used to reach a given point of this region? To answer these questions, we associate a long-term \textit{dynamical} version with the assumed one-shot game. The dynamical game we consider can be put under strategic-form. The set of players is $\mathcal{I}=\{1,2\}$. The strategies or power control policies are defined as follows:
\begin{equation}
    \begin{array}{lllccc}
     \sigma_{i,t}& :& & \mathcal{S}_i^{t} & \to & \mathcal{A}_{i}\\
 & & &(s_{i}(1),\ldots,s_{i}(t)) & \mapsto & a_i(t)\\
\end{array}
\end{equation}
where $t\geq 1$ is the stage or block index, $\mathcal{S}_i^{t}=\mathcal{S}_i^{1}\times \mathcal{S}_i^{2}\times \ldots \times \mathcal{S}_i^{t}$, $\mathcal{S}_i$ being the set of signals or observations of Node $i$. The observation $s_i(t) \in \mathcal{S}_i$ corresponds to the image Node $i$ has about the global channel state $x_0(t)$ at stage $t$. This image is assumed to be the output of a memoryless channel whose conditional probability is $\daleth_{i}(s_{i}|x_{0})$. The long-term utility of Node $i$ is defined as:
\begin{equation}\label{eq9}
U_i(\sigma_1,\sigma_2)=\mathbb{E}\left[\frac{1}{T}\sum_{t=1}^{T} u_i(x_{0}(t),a_{1}(t),a_{2}(t))\right].
\end{equation}


\section{Utility region characterization and proposed power control policies}\label{sec:utility-region}

It is important to know that the characterization of the feasible utility region of dynamic games (which includes repeated games as a special case) with an arbitrary observation structure is still an open problem  \cite{maschler}. Remarkably, as shown recently in \cite{larrousse}\cite{benja}, the problem can be solved for some setups which are quite generic in wireless communications. Indeed, when $T$ is assumed to be large, the random process associated with the system state $X_0(1), ...,X_0(T)$ is i.i.d., and the observation structure given by $\daleth_1,\daleth_2$, it is possible to derive a coding theorem which characterizes the joint probability distribution $Q(x_0,a_1,a_2)$ which can be attained. This is precisely the corresponding theorem we propose to exploit here to determine numerically the feasible utility region of the considered dynamic game and to derive a good power control policy. After \cite{larrousse}\cite{larrousse-tit-2015} a joint distribution $Q(x_0,a_1,a_2)$  is achievable in the limit $T\rightarrow\infty$ if and only if it factorizes as:
\begin{multline}\label{achie}
   Q(x_0,a_1,a_2)=\sum_{V,s_{1},s_{2}}\rho(x_{0})P_{V}(v)\times \daleth(s_{1},s_{2}|x_{0})\times \\
     P_{A_{1}|S_1, V}(a_{1}|s_1, v) P_{A_{2}|S_2, V}(a_{2}|s_2, v)
    \end{multline}
where: $\rho$ is the probability distribution of the channel state; $\daleth$ is the joint conditional probability which defines the assumed observation structure; $V \in \mathcal{V}$ is an auxiliary random variable or lottery which can be proved to improve the performance in general (see \cite{benja} for more details). By exploiting this result, it comes that a pair of expected payoffs $(U_1, U_2)$ is achievable if it writes as $(\mathbb{E}_Q(u_1), \mathbb{E}_Q(u_2))$ where $Q$ factorizes as (\ref{achie}). By using a time-sharing argument, the achievable utility region has to be convex. Therefore, the Pareto frontier of the utility region can be obtained by maximizing the weighted utility
\begin{equation}\label{eq:wlambda}
W_{\lambda} = \lambda \mathbb{E}_Q(u_1) + (1-\lambda) \mathbb{E}_Q(u_2)
\end{equation}
 with respect to $Q$, for every $\lambda \in [0,1]$. In (\ref{achie}), $\rho$ and $\daleth$ are given. Thus, $W_{\lambda}$ has to be maximized with respect to the triplet $(P_{A_{1}|S_1, V}, P_{A_{2}|S_2, V}, P_{V})$. In this paper, we restrict our attention to the optimization of  $(P_{A_{1}|S_1, V}, P_{A_{2}|S_2}$ for a fixed lottery $P_{V})$ and leave the general case as an extension. Given this, the optimization problem at hand becomes a bilinear problem: the function to be maximized is bilinear and the optimization space is the unit simplex $\Delta(\mathcal{A}_1 \times \mathcal{A}_2 \times \mathcal{V})$. It can be checked (see e.g., \cite{Gallo}) that the optimal solutions of this problem lies on the vertices of the unit simplex $\Delta(\mathcal{A}_1 \times \mathcal{A}_2 \times \mathcal{V})$. This key observation indicates that there is no loss of optimality by only searching for functions instead of a general conditional probability distribution. The variables to be optimized are thus functions which we denote by $f_{i}: \mathcal{S}_i\times \mathcal{V}  \rightarrow \mathcal{A}_{i}$. The corresponding bilinear program can be solved by using techniques such as the one proposed in \cite{Konno}, but global convergence is not guaranteed. Two other relevant numerical techniques have also been proposed in \cite{Gallo}. The first technique is based on cutting plane while the second one consists of an enlarging polytope approach. For both techniques, convergence may also be an issue since for the first technique, no convergence result is provided and for the second technique, cycles may appear \cite{VaShe}. To solve the convergence issue, we propose to exploit the sequential best-response dynamics (see e.g., \cite{fudenberg-book-1991}, \cite{sam}), which has been used recently for the interference channel \cite{achal}. The sequential best-response dynamics applied in the considered setup works as follows. At iteration $0$, an initial choice for the conditional probability which defines the decision of the two nodes is made: $(P_{A_{1}|S_1, V}, P_{A_{2}|S_2, V}) =  (P_{A_{1}|S_1, V}^{(0)}, P_{A_{2}|S_2, V}^{(0)})$. At iteration $1$, (\ref{eq:wlambda}) is maximized with respect to $P_{A_{1}|S_1, V}^{(1)}$. At iteration $2$,
 $P_{A_{2}|S_2, V}$ is updated by maximizing (\ref{eq:wlambda})  with respect to $P_{A_{2}|S_2, V}^{(2)}$, and so on. Since at each iteration, the same utility (namely, $W_{\lambda}$) is maximized, the sequential best-response dynamics is guaranteed to converge; the proof follows by induction or using a more general argument such as exact potentiality. Global convergence is however not guaranteed in general. The only argument the authors can provide is that exhaustive simulations show that the loss of optimality induced by operating at the points the algorithm reaches is small. An important point to note here is that the proposed procedure provides stationary power control strategies or power control decision functions, which can be used in practice.

\section{Numerical performance analysis}
\label{sec:numerical-perf-analysis}

By default, the following simulation setup is assumed; otherwise, the parameters are explicitly mentioned in the figures. Set of possible power levels $\forall i \{1,2\}, \mathcal{P}_i =\mathcal{P}_i' $: $M=25$, $P_i^{\min}(\mathrm{dB}) = -20$, $P_i^{\max}(\mathrm{dB}) = +20$, and the power increment in dB equals $\frac{40}{24}$. Set of possible channel gains: $\forall i \{1,2\}, \mathcal{G}_i =\mathcal{G}_i' $: $N=20$,
$g^{\min}_{i} = 0.01$, $g^{\max}_{i} = 10$, and the channel gain increment equals $\frac{10-0.01}{19}$. The different means of the channel gains are given by: $(\bar{g}_1,\bar{g'}_1,\bar{g}_2,\bar{g'}_2)=(1,1.9,1,1.9)$. The communication efficiency function is chosen as in \cite{belmega-tsp-2011}:  $\varphi(x) = e^{-\frac{c}{x}}$ with $c=2^r -1$, $r$ being the spectral efficiency in bit/s per Hz  \cite{belmega-tsp-2011}. In the simulations provided we have either $r=1$ or $r=3$, which are reasonable values for wireless sensor network communications.

\begin{figure}[htbp]
  \includegraphics[width=8.5cm,height=6cm]{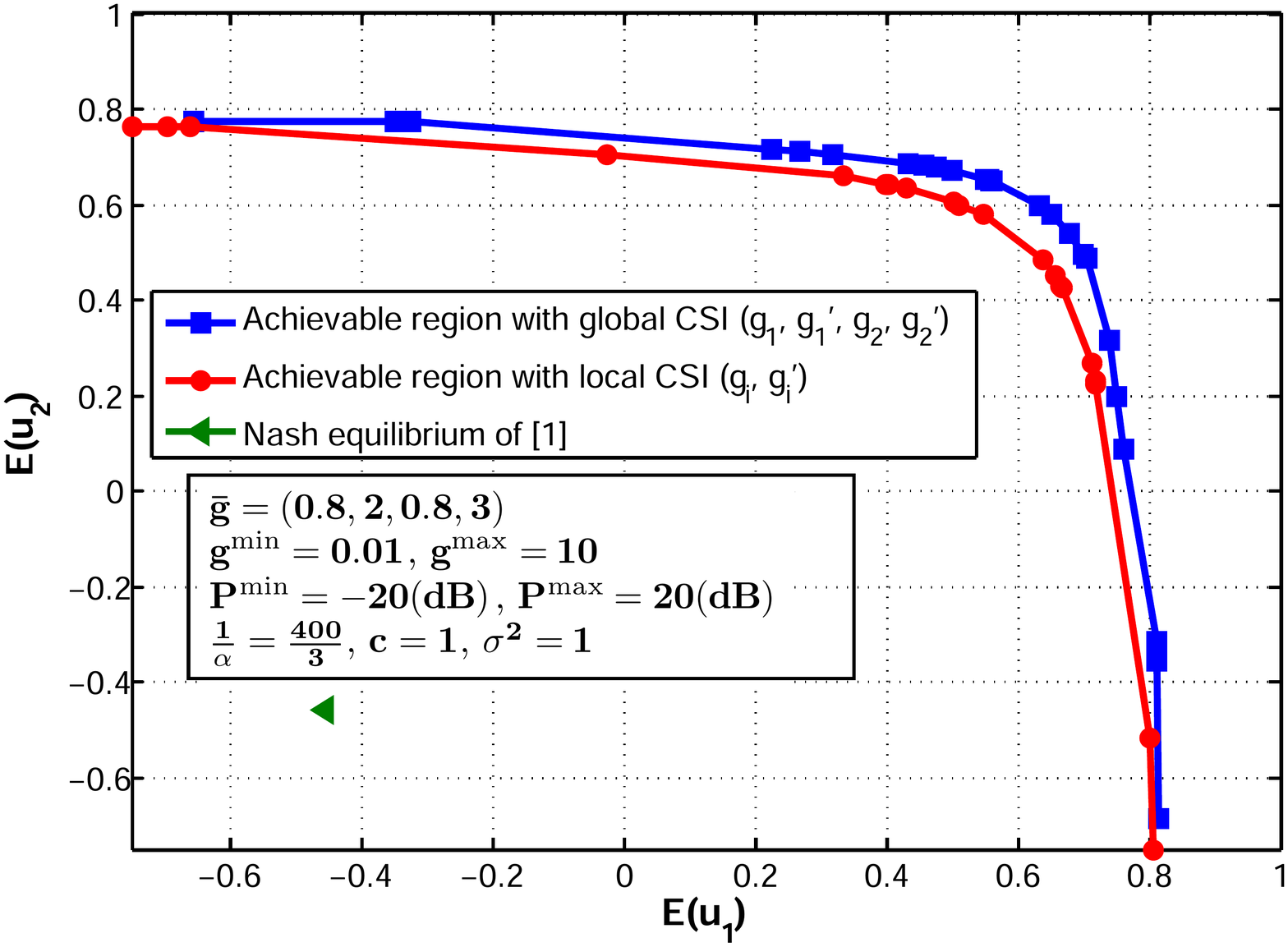}
\caption{Achievable utility region for two scenarios of partial information for the source nodes: global CSI and local CSI. The performance can be compared to the isolated point which represents the one-shot forwarder's dilemma Nash equilibrium \cite{hub}.}
\label{fig2}
\end{figure}

Fig. \ref{fig2}  represents the long-term utility region of the dynamical game presented in this paper. It has been obtained by applying the sequential best-response dynamics to (\ref{eq:wlambda}) for different values of $\lambda$. Three scenarios are considered. The point in the bottom left corner corresponds to the performance of the Nash equilibrium of the forwarder's dilemma of \cite{hub}. The two curves corresponds to achievable pair of long-term utilities when global channel state information is known from the two source nodes and when they have only local channel state information. First, it is seen that our new formulation provides very significant improvements over the approach which consists in studying the packet forwarding problem without accounting for the link quality fluctuations. Second, the framework we use in this paper allows one to quantify what is gained when more information. Here, in particular, it is seen that global channel state information only provide a relatively small gain for the utilities. This therefore advocates the use of distributed power control policies which only rely on local channel state information.

\begin{figure}[htbp]
  \includegraphics[width=8.5cm,height=6cm]{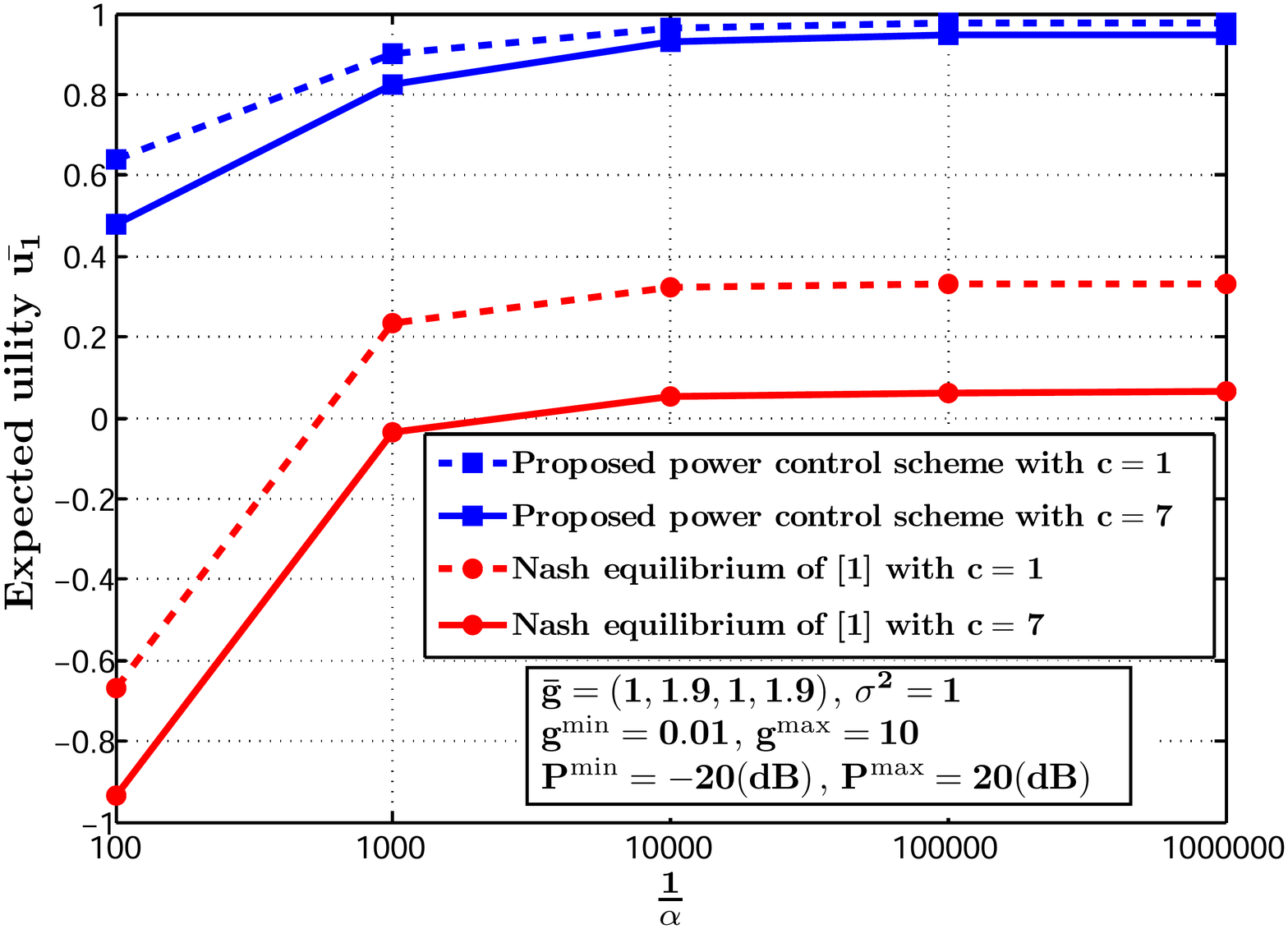}
\caption{Long-term utility with local CSI against the reciprocal of the weight assigned to the energy cost part in (1) that is, $\frac{1}{\alpha}$.}
\label{fig3}
\end{figure}

Fig. \ref{fig3} represents the long-term utility of Node $1$ (which coincides with the one of Node $2$ since considered scenarios are symmetric) against the reciprocal of the weight assigned to the energy consumption namely, $\frac{1}{\alpha}$; two values for the spectral efficiency are retained $r=1$ bit/s per Hz and $r=3$ bit/s per Hz. The figure clearly shows a very significant gain in terms of utility and fully supports the proposed approach i.e, the power control policies obtained through the proposed numerical procedure outperform classical policies.

\begin{figure}[htbp]
  \includegraphics[width=8.5cm,height=6cm]{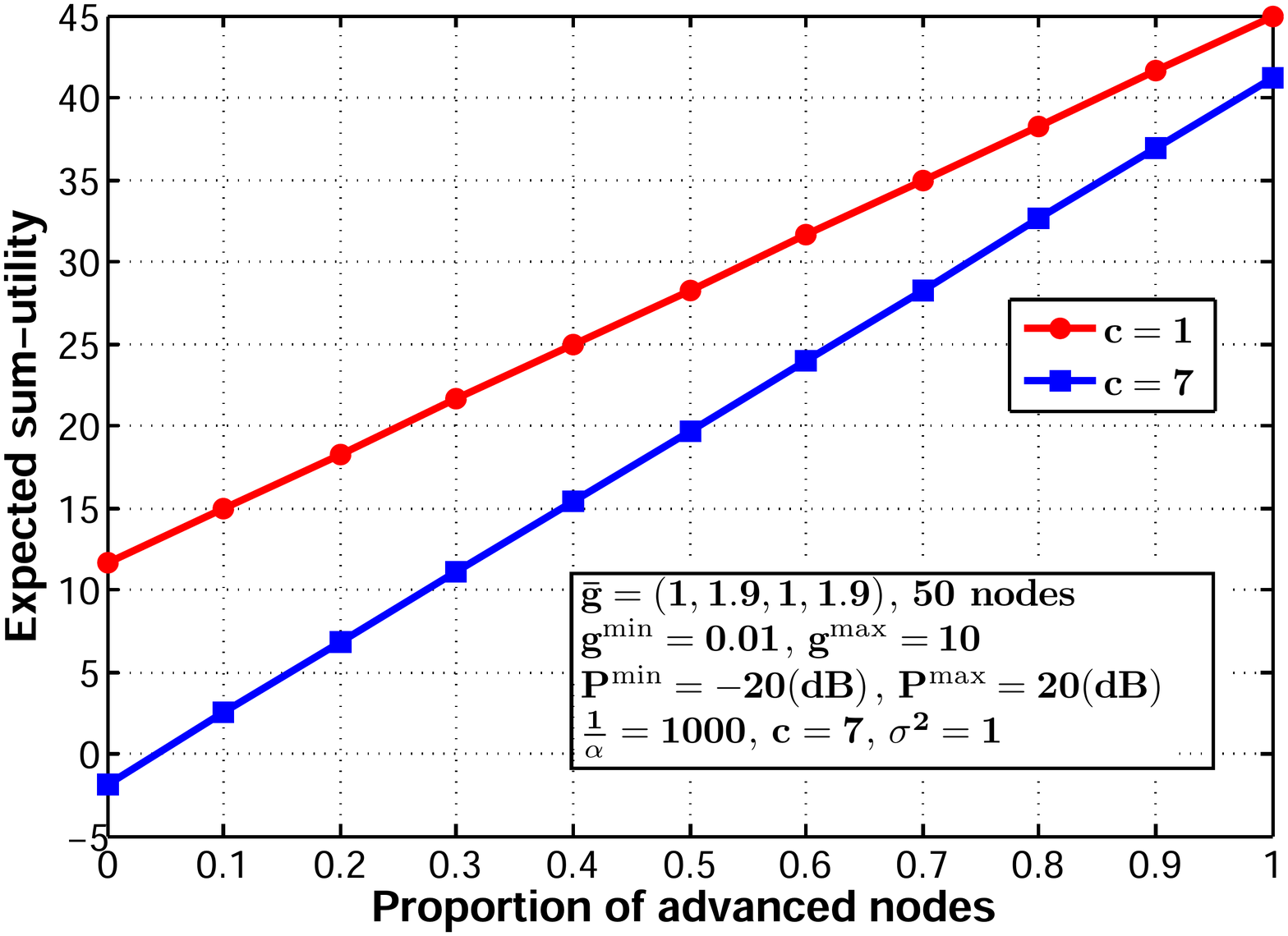}
\caption{Expected sum-utility against the proportion of advanced nodes for an ad hoc network of $50$ nodes with $4-$node interactions; by "advanced" it is meant that the nodes implement the power control policy proposed in this work while the other nodes implement the one-shot forwarder's dilemma Nash equilibrium power control policies.}
\label{fig4}
\end{figure}

Fig. \ref{fig4} corresponds to an ad hoc network which comprises $50$ nodes and $4$-node interactions are assumed, that are not all disjoint.  i.e, there may exist two sets of $4-$node containing at most three same nodes. The figure shows the performance gain against the fraction of nodes which implement the proposed power control policies while the other nodes implement the Nash equilibrium forwarder's dilemma policies of \cite{hub}. The performance gains scales linearly with the fraction of "advanced nodes" and the network performance is very significantly improved when one compares the case where the fraction equals $1$ to the one where it equals $0$.

\section{Conclusion}\label{sec6}

One of the contributions of this work is to generalize the famous and insightful model of forwarder's dilemma \cite{hub} by accounting for channel gain fluctuations. The problem of knowledge about global CSI therefore appears. We have seen that it is possible to characterize the performance of the studied system even in the presence of partial information; the corresponding observation structure is arbitrary provided the observations are generated by a discrete memoryless channel denoted by $\daleth$ in this paper. In terms of performance, designing power control policies which exploit as well as possible the available knowledge is shown to lead to very significant gains. A very significant extension of the present work would be to relax the i.i.d. assumption on the system state. In this work, the system state corresponds to the global channel state and the i.i.d. assumption is known to be very reasonable but, in other setups, where the state represents e.g., a queue length or an energy level, the used framework needs to be extended.

\end{document}